\documentclass{nic-series}
\usepackage[pdftex]{color}

\usepackage{epsf}

\def\jnl@style#1{{\rmfamily#1}}%
\def\jref@jnl#1{{\jnl@style#1}}%

\newcommand\aj{\jref@jnl{AJ}}%
\newcommand\araa{\jref@jnl{ARA\&A}}%
\newcommand\apj{\jref@jnl{ApJ}}%
\newcommand\apjl{\jref@jnl{ApJ}}%
\newcommand\apjs{\jref@jnl{ApJS}}%
\newcommand\ao{\jref@jnl{Appl.~Opt.}}%
\newcommand\apss{\jref@jnl{Ap\&SS}}%
\newcommand\aap{\jref@jnl{A\&A}}%
\newcommand\aapr{\jref@jnl{A\&A~Rev.}}%
\newcommand\aaps{\jref@jnl{A\&AS}}%
\newcommand\azh{\jref@jnl{AZh}}%
\newcommand\baas{\jref@jnl{BAAS}}%
\newcommand\jcop{\jref@jnl{J.~Comp.~Phys.}}%
\newcommand\jkas{\jref@jnl{Journ.~Korean.~Astron.~Soc.}}%
\newcommand\jrasc{\jref@jnl{JRASC}}%
\newcommand\memras{\jref@jnl{MmRAS}}%
\newcommand\mnras{\jref@jnl{MNRAS}}%
\newcommand\na{\jref@jnl{New Astron.}}%
\newcommand\nar{\jref@jnl{New Astron. Rev.}}%
\newcommand\pra{\jref@jnl{Phys.~Rev.~A}}%
\newcommand\prb{\jref@jnl{Phys.~Rev.~B}}%
\newcommand\prc{\jref@jnl{Phys.~Rev.~C}}%
\newcommand\prd{\jref@jnl{Phys.~Rev.~D}}%
\newcommand\pre{\jref@jnl{Phys.~Rev.~E}}%
\newcommand\prl{\jref@jnl{Phys.~Rev.~Lett.}}%
\newcommand\pasa{\jref@jnl{PASA}}%
\newcommand\pasp{\jref@jnl{PASP}}%
\newcommand\pasj{\jref@jnl{PASJ}}%
\newcommand\qjras{\jref@jnl{QJRAS}}%
\newcommand\skytel{\jref@jnl{S\&T}}%
\newcommand\solphys{\jref@jnl{Sol.~Phys.}}%
\newcommand\sovast{\jref@jnl{Soviet~Ast.}}%
\newcommand\ssr{\jref@jnl{Space~Sci.~Rev.}}%
\newcommand\zap{\jref@jnl{ZAp}}%
\newcommand\nat{\jref@jnl{Nature}}%
\newcommand\iaucirc{\jref@jnl{IAU~Circ.}}%
\newcommand\aplett{\jref@jnl{Astrophys.~Lett.}}%
\newcommand\apspr{\jref@jnl{Astrophys.~Space~Phys.~Res.}}%
\newcommand\bain{\jref@jnl{Bull.~Astron.~Inst.~Netherlands}}%
\newcommand\fcp{\jref@jnl{Fund.~Cosmic~Phys.}}%
\newcommand\gca{\jref@jnl{Geochim.~Cosmochim.~Acta}}%
\newcommand\grl{\jref@jnl{Geophys.~Res.~Lett.}}%
\newcommand\jcp{\jref@jnl{J.~Chem.~Phys.}}%
\newcommand\jgr{\jref@jnl{J.~Geophys.~Res.}}%
\newcommand\jqsrt{\jref@jnl{J.~Quant.~Spec.~Radiat.~Transf.}}%
\newcommand\memsai{\jref@jnl{Mem.~Soc.~Astron.~Italiana}}%
\newcommand\nphysa{\jref@jnl{Nucl.~Phys.~A}}%
\newcommand\physrep{\jref@jnl{Phys.~Rep.}}%
\newcommand\physscr{\jref@jnl{Phys.~Scr}}%
\newcommand\planss{\jref@jnl{Planet.~Space~Sci.}}%
\newcommand\procspie{\jref@jnl{Proc.~SPIE}}%
\newcommand\znat{\jref@jnl{Z.~Naturforsch}}%
\newcommand{\sm}[1]{\rm{{\scriptsize #1}}}
\newcommand{\simle} {\,{}^<_{\sim}\,}
\newcommand{\simge} {\,{}^>_{\sim}\,}

\newcommand{\cm}{\rm{cm}}

\newcommand{\pc}{\rm{pc}}

\newcommand{\Myr}{\mbox{Myr}}
\newcommand{\muG}{\mu\mbox{G}}

\newcommand{\KB}{{K\"ortgen \& Banerjee (2015)}}
\newcommand{\MS}{{Mestel \& Spitzer}}

\newcommand{\bef}{\begin{figure}[!t]}
\newcommand{\eef}{\end{figure}}

\def\figtwo@scaling{0.48}

\def\showtwo#1#2{
  \centering
  \leavevmode
  \includegraphics[width=\figtwo@scaling\linewidth]{#1.pdf}
  \includegraphics[width=\figtwo@scaling\linewidth]{#2.pdf}
}

\def\figthree@scaling{0.30}

\def\showthree#1#2#3{
  \centering
  \leavevmode
  \includegraphics[width=\figthree@scaling\linewidth]{#1.pdf}
  \includegraphics[width=\figthree@scaling\linewidth]{#2.pdf}
  \includegraphics[width=\figthree@scaling\linewidth]{#3.pdf}
}

\begin{document} 

\title{Formation of star-forming clouds from the magnetised, diffuse
  interstellar medium}

\author{Robi Banerjee \& Bastian K\"ortgen}

\institute{Hamburger Sternwarte, Universit\"at Hamburg, \\
         Gojenbergsweg 112, 21029 Hamburg, Germany
%\\
%         \email{banerjee@hs.uni-hamburg.de}
}

\maketitle


\begin{abstracts}
  Molecular clouds, the birthplaces of stars in galaxies, form
  dynamically from the diffuse atomic gas of the interstellar medium
  (ISM). The ISM is also threaded by magnetic fields which have a
  large impact on its dynamics. In particular, star forming regions
  must be magnetically {\em supercritical} in order to accomodate gas
  clumps which can collapse under their own weight.  Based on a
  parameter study of three dimensional magneto-hydrodyamical (MHD)
  simulations, we show that the long-standing problem of how such
  supercritical regions are generated is still an open issue.
\end{abstracts}

\section{Introduction}

Present day stars form within the densest regions of molecular clouds
(MCs) and giant molecular clouds (GMCs), in gravitationally unstable
cores and clumps.
% which often reside at the junctions of filaments\cite{AndrePPVI14}.
Our common understanding is that those MCs and GMCs form from the
diffuse, atomic (HI) gas within timescales of less than $10\,\Myr$
\cite{WilliamsPPIV00, Blitz07}. The generation of filaments and
substructures within GMCs is primarily controlled by magnetic fields
and turbulence \cite{Shu87,MacLow04,AndrePPVI14}.  In particular,
magnetic fields are an elemental part of the interstellar medium
\cite{Beck12, Crutcher12, Heiles12, Beck13, LiHBPPVI14, LiPPVI14}
which have a large impact on the dynamics of the ISM on various
spatial scales \cite{Li11Nature, LiHB15, Pillai15, Franco15} as the
magnetic energy density is comparable to the thermal energy density of
the ISM\cite{Heiles05}.

One long-standing issue is the formation of supercritical clumps and
cores. Similarly to thermal pressure, magnetic fields prevent
contraction of otherwise (thermally) self-gravitating gas clumps if
the magnetic fields are strong enough. Therefore, gaseous
overdensities must be magnetically {\em supercritical}, quantified by
the mass-to-flux ratio, $\mu$, to collapse and to subsequently allow
the formation of stars.

Already in 1956, \MS\, realised that molecular clouds should be
magnetically {\em subcritical} assuming field strengths that
correspond to the equipartition of magnetic and kinetic energy density
within the ISM\cite{Mestel56}. To generate supercritical cloud cores
out of those subcritical conditions, they suggested that the
non-perfect coupling between charged particles and neutrals, i.e. the
ambipolar diffusion (AD) drift, could locally reduce the mass-to-flux
ratio which allows the cloud to break up and to form stars. For a long
time this was the standard theory of star formation out of the
magnetised ISM \cite{Shu87, Mouschovias99}. In this fairly static
``standard model'' of magnetically-supported, AD-mediated
supercritical cores, low-mass stars would form by the the slow
gravitational contraction of isolated cores containing a very small
fraction of the clouds' mass.  This picture would also account for the
very low observed global star formation efficiency (SFE) of giant
molecular clouds\cite{Myers86, Evans09}. The slow contraction results
from the typical timescale for ambipolar diffusion, $t_{\sm{AD}}$,
which is an order of magnitude larger than the free-fall time,
$t_{\sm{ff}}$, of individual cloud cores (their ratio is about
$t_{\sm{AD}}/t_{\sm{ff}} \approx 10\,(x_e/10^{-7})$, where $x_e$ is
the ionisation fraction). On the other hand, rather recent
observations by Crutcher (2009)\cite{Crutcher09} of individual cloud
cores including Zeeman measurements to determine their magnetic field
distribution indicate that idealised models of ambipolar-diffusion
driven star formation are unlikely to be operative. In idealised
models with ordered background magnetic fields, efficient ambipolar
diffusion would lead to a local increase of the mass-to-flux ratio
towards the centre of cloud cores which is not seen in their observed
sample\cite{Crutcher09} (but see also Bertram et
al. (2012)\cite{Bertram12} on the difficulty to interpret those
observations).
%% from Crutcher et al. 2009 Our experiment is limited to four clouds,
%% and we can only directly test the predictions of the extreme-case
%% "idealized" models of ambipolar-diffusion driven star formation,
%% which have a regular magnetic field morphology. Nonetheless, our
%% experimental results are not consistent with the "idealized" strong
%% field, ambipolar diffusion theory of star formation.

However, present-day models of star formation also account for the
fact that molecular clouds are also pervaded by supersonic random
motions, i.e. turbulence \cite{Zuckerman74, Larson81,
  Solomon87}. Eventually, this resulted in a paradigm shift of the
theory of star formation where magnetic fields only play a minor role
and supersonic, super-Alfv\'enic turbulence controls the star
formation efficiency within molecular clouds \cite{MacLow04,
  Elmegreen04, Ballesteros07, DobbsPPVI14, PadoanPPVI14}. As a
consequence, the magnetic fields are expected to be highly disordered
rather than being an ordered background field. Hence, idealised models
of ambipolar diffusion drift should not apply \cite{Crutcher10,
  Crutcher10b}. Additionally, the AD characteristic timescale is
expected to decrease in this case
%, suggestingthat it may be comparable to the free-fall time
\cite{Fatuzzo02, Heitsch04, Kudoh07, Kudoh11} and other
diffusive effects like turbulent reconnection might be operative
\cite{Vishniac99, Eyink13}.
% Thus, the gravitational contraction and collapse of a star-forming
% region could occur rapidly, essentially on timescales corresponding
% to those of a magnetically supercritical region, which is of the
% order of the free-fall time \citep{OGS99,  Heitsch01, Vazquez05,
% GalvanMadrid07}. 
Indeed, a number of studies have suggested that both MCs
\cite{McKee89} and their clumps \cite{Myers88, Bertoldi92,
  Crutcher99, Bourke01, Crutcher03, Troland08} are close to being
magnetically critical, with a moderate preference for being
supercritical. Moreover, recent compilations of observational
data show that cloud cores and clumps with column densities of $N
\simge 2\times 10^{21}\,\cm^{-2}$ are essentially {\em all}
supercritical (see Fig.~\ref{fig:crutcher12-fig7}).
%%This implies that, if a MC is subcritical, it is expected to be only
%%moderately so as well, in which case the time for cores within it to
%%become locally supercritical may be almost as short as the cores' free-fall time \citep{CB01, VS_etal05}. 

\bef
  \centering
  \leavevmode
  \includegraphics[width=.7\linewidth]{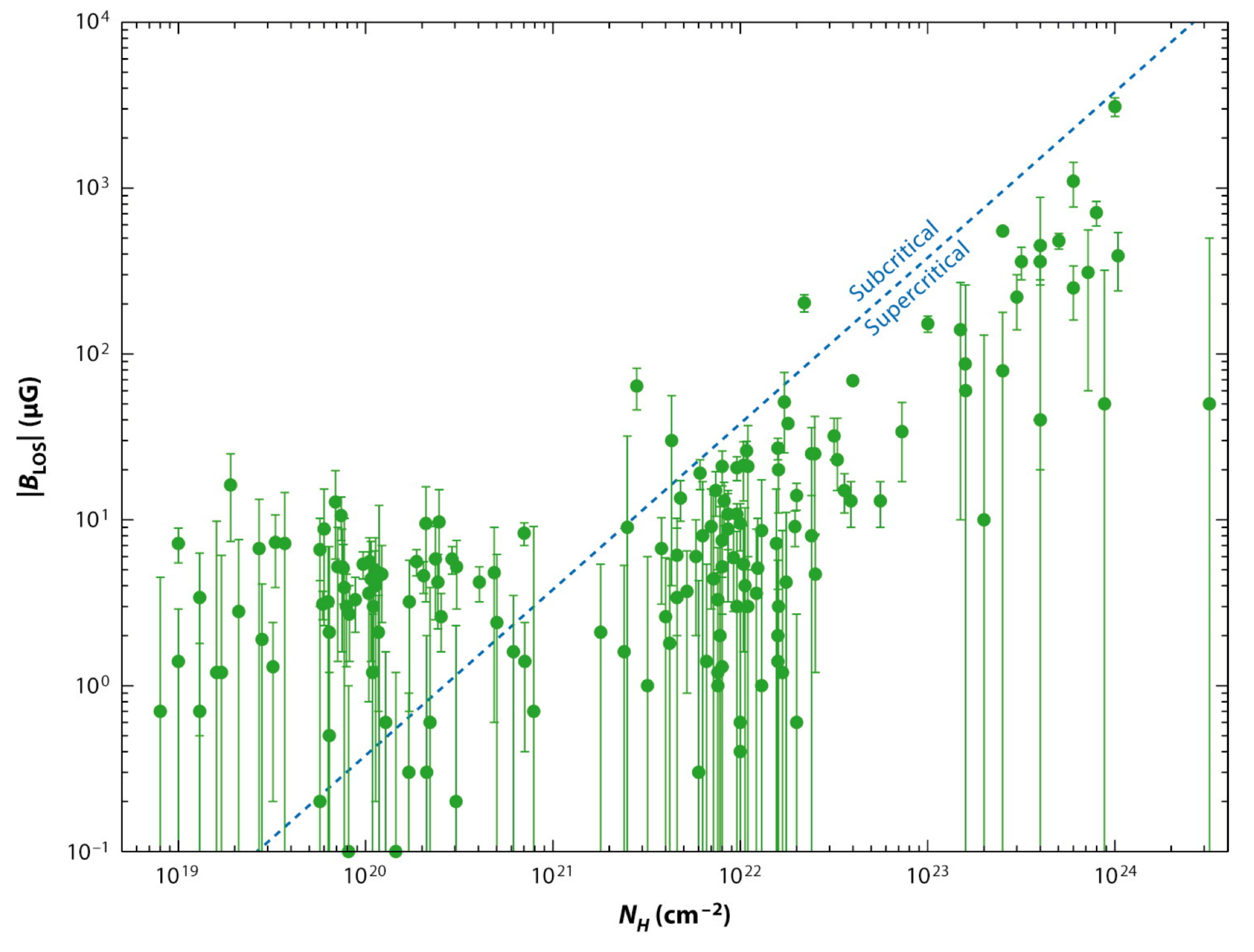}
  \caption{These observational data summarise the main motivation for our proposed
    study: How do subcritical (HI) clouds become supercritical (H$_2$) clouds? Our
    previous studies have shown that it is everything but trivial to
    build up supercritical clouds out of the magnetised interstellar medium,
    because the mass-to-flux ratio is fairly well conserved,
    even in the presence of ambipolar diffusion and enhanced non-ideal
    MHD setups \cite{Vazquez11a, Koertgen15}.
    \label{fig:crutcher12-fig7} 
    [From Crutcher (2012)\cite{Crutcher12}]}
\eef

Whether those supercritical cloud cores and clumps are the result of
ambipolar diffusion together with random motions in the ISM is far
from being certain and has to be investigated further. For instance,
recently Heitsch \& Hartmann (2014)\cite{Heitsch14} argued in their
parameter study that ambipolar diffusion in concert with turbulence is
unlikely to control the formation of supercritical cores and hence
star formation. They again propose an alternative scenario where large
scale
%, initially supercritical, 
flows are the main driver to generate supercritical cores. This idea,
where supercritical clouds could be assembled from large scale flows
was already discussed in \MS\, (1956)\cite{Mestel56} as an
alternative to the AD-mediated scenario and to avoid the ``magnetic
flux problem''. But only in combination with supersonic turbulence
this scenario becomes more feasible because gravitational
fragmentation could be suppressed during the assembly of the clouds by
those turbulent motions \cite{Passot95, Hartmann01}. This accumulation
idea would also support a number of recent observations which show
that magnetic fields are dynamically important on all scales in the
Milky Way and other spiral galaxies \cite{Fletcher11, Li11Nature,
  Pillai15, PLANCK-2015-XXXV, LiHB15, Franco15}. This is particular
evident from Fig.~1: The low column density HI gas is magnetically
{\em sub}critical, whereas clouds which exceed columns of
$N \simge 2\times 10^{21}\,\cm^{-2}$ are magnetically {\em
  super}critical.

In the presented numerical parameter study, we investigated the
possibility of diffusion mediated generation of {\em supercritical}
clouds showing that it is unlikely that such unstable clouds can be
build up from {\em subcritical} HI-clouds.  

\section{Numerical method and initial conditions}

\bef
\centering
\showtwo{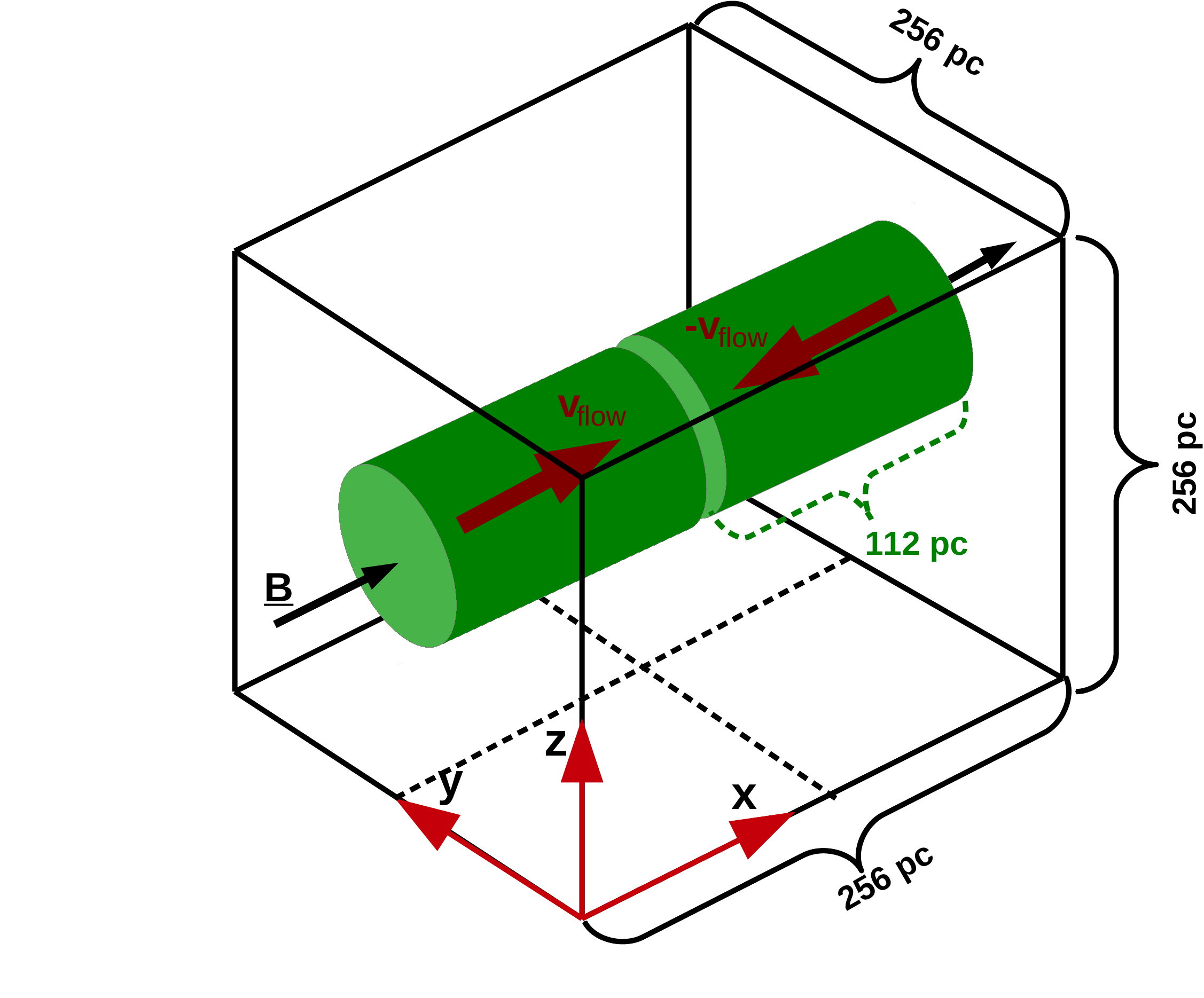}
               {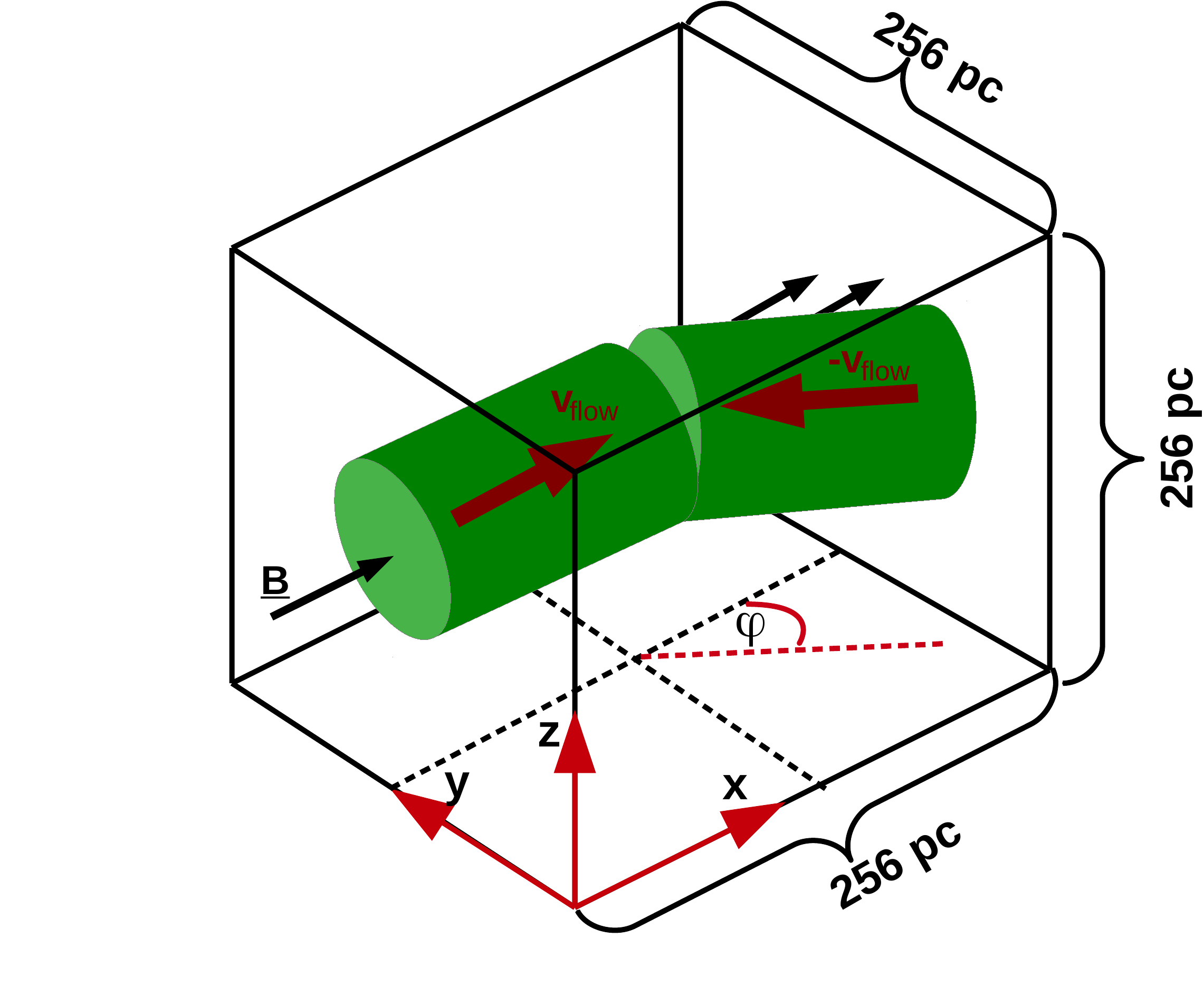}
\caption{Setup of the initial conditions. {\it Left}: The well studied
  head-on collision where both flows are in the direction of the
  background magnetic field. {\it Right}: Flows with an inclination
  angle with respect to the background magnetic field. This setup
  resembles the impact of external driving of such flows (e.g. driving
  by SN blast waves). [From \KB\cite{Koertgen15}]}
  \label{fig:setup}
\eef

For these studies we used the FLASH adaptive mesh refinement (AMR)
code \cite{FLASH00, Dubey08}. In addition to the basic ideal MHD
equations (for which we employ the Bouchut solver\cite{Bouchut07,
  Bouchut10, Waagan10}) we also used the {\em ambipolar diffusion}
module developed from Duffin \& Pudritz
(2008)\cite{Duffin08}. Additionally, self-gravity as well as heating
and cooling processes were included in those simulations. For the
latter, we followed the treatment by Koyama \& Inutsuka
(2002)\cite{Koyama02} (an analytic simplification of their detailed
calculation in \cite{Koyama00, Vazquez07, Banerjee09a}). To capture
the build-up of self-gravitating cores within the molecular clouds we
used {\em sink particles}\cite{Federrath10} in addition to the Jeans
refinement criterion (i.e. the Truelove
criterion\cite{Truelove97}). In particular the detailed sink particle
approach allows us to unambiguously identify supercritical, collapsing
regions which are importent for our studies quantifying the star
formation ability from the magnetised ISM.

Our initial setups for those studies are similar to the ones described
in \cite{Vazquez07, Banerjee09a, Vazquez11a} (see also
Fig.~\ref{fig:setup}) where the build-up of molecular clouds is
modeled by the collision of cylindrical streams of warm neutral HI gas
(WNM).  Each flow is $l=112\,\pc$ long and has a radius of
$r=64\,\pc$. The bulk flows are slightly supersonic with typical Mach
numbers of $\mathcal{M}_\mathrm{f}=2$. On top of those bulk motions, a
turbulent velocity field is superimposed which triggers initial
instabilities like the non-linear thin-shell instability (NTSI)
\cite{Vishniac94} and subsequently leads to fragmentation of the
cloud. The initially uniform magnetic field has a strength of
$B =\left\{3,4,5\right\}\,\muG$ corresponding to mass--to--flux ratios
of $\mu/\mu_{\sm{crit}} \approx 1, 0.7, 0.6$ if the critical value
$\mu_{\sm{crit}} \approx 0.13/\sqrt{G}$ is
applied\cite{Mouschovias76}.

Furthermore, we also studied the impact of an oblique angle of the
flows with respect to the background magnetic field (see right panel
of Fig.~\ref{fig:setup}). Those oblique flows are more realistic than
the head-on flows and could be generated, for instance, by supernova
shock waves and by the gravitational potential of spiral arms. The
motion of the flow at an inclination with respect to the magnetic
field results in enhanced magnetic diffusivity (by numerical
diffusion\cite{Heitsch05,Heitsch08b,Inoue08}). Again, those flows
resemble streams of the WNM in a thermally bistable configuration
\cite{Field65}. The flows are studied with different oblique angles,
which are varied from 10$^{\rm o}$ to 60$^{\rm o}$, different initial
magnetic fields strengths and different strength of the initial
turbulence ranging from subsonic to supersonic velocity
fluctuations. For details on the numerical setup and our initial
conditions see \KB\cite{Koertgen15}.

\section{Results}

\bef
\centering
  \includegraphics[width=0.45\linewidth]{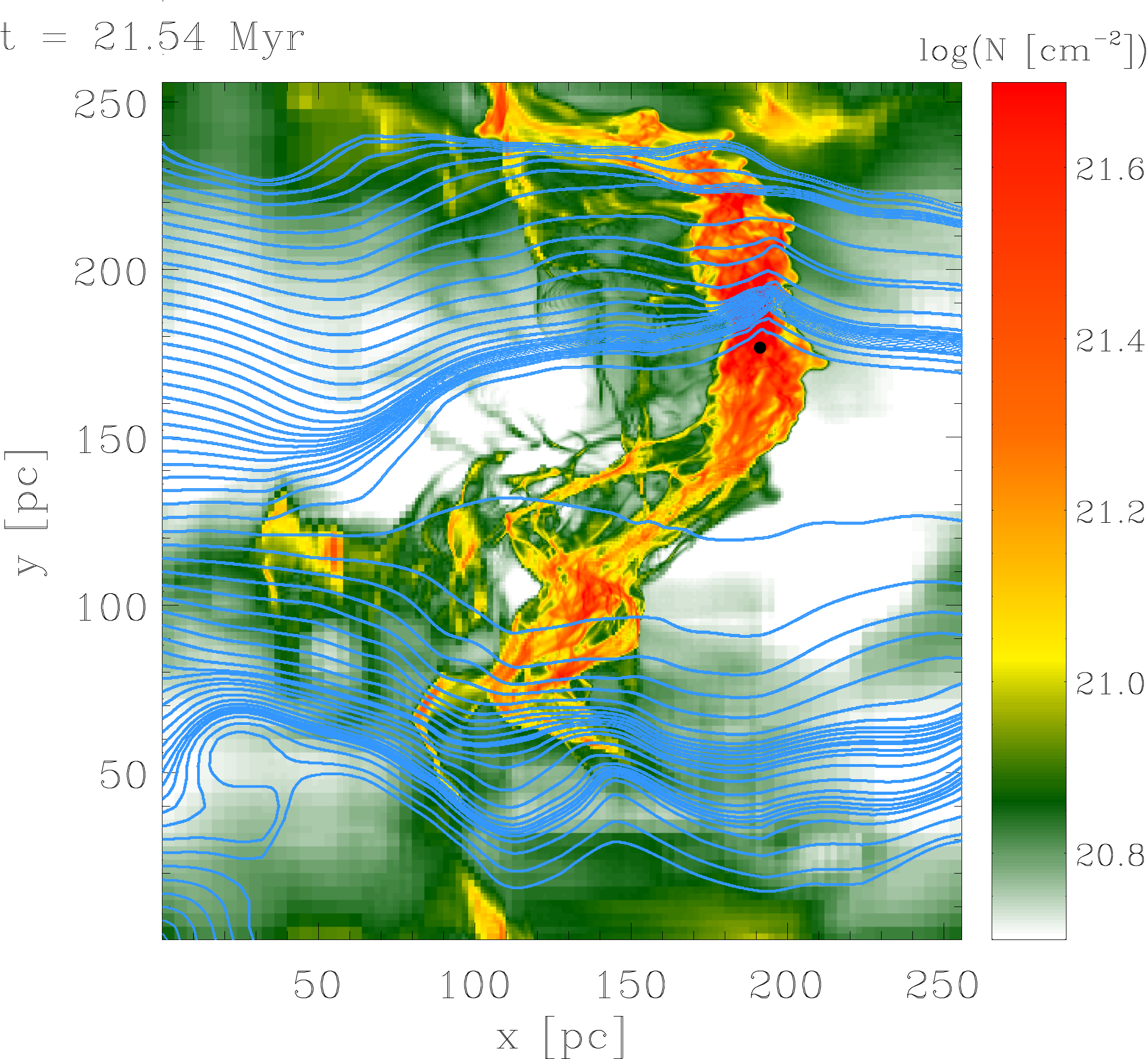}
  \includegraphics[width=0.45\linewidth]{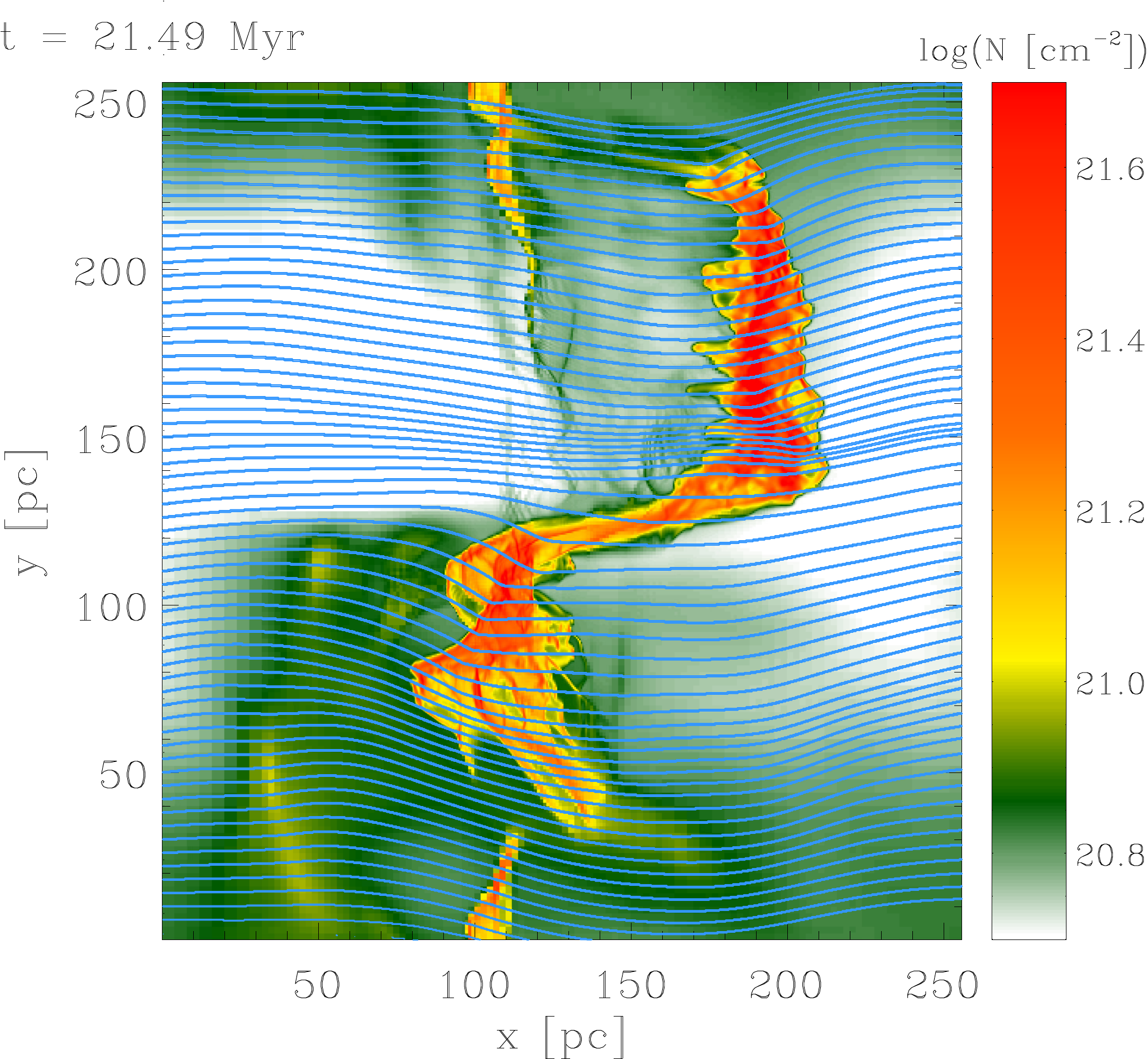}
  \caption{Results from colliding flow simulations
    investigating the formation of molecular
    clouds. Here, the flows collide with an oblique angle of
    $60^{\circ}$. {\em Left panel}: The weak field case ($3\,\muG$).
    The {\em Right panel} shows the same
    situations in the case of a stronger background
    magnetic field of $5\,\muG$. In the case of a weak
    magnetic field supercritical cloud cores can form
    that allow the formation of stars (marked with black
    dots). Stronger initial magnetic fields prohibit
    the formation of stars even in the case of large
    oblique angles of the flows. The blue stream-lines
    indicate the magnetic field morphology in the
    projected 2D plane. [From \KB\cite{Koertgen15}]
  \label{fig:mc-form-tilted}}
\eef

As can be seen from Fig.~\ref{fig:mc-form-tilted}, the different
initial field strengths have significant implications for the
resulting dynamical behaviour of the molecular cloud. The main
difference comes about in efficiency to form stars (or not). In the
case of a rather weak background field of $3\,\muG$ supercritical star
forming clumps can be generated whereas in the case of a slightly
stronger, but more realistic, magnetic field star formation is fully
suppressed. Note that, due to the oblique flows with an angle of
$60^{\circ}$ the effective mass-to-flux ratios are $0.73$ in the
$3\,\muG$ case and $0.44$ in the $5\,\muG$ case. That means that both
cases are initially sub-critical, but only in the cases of the weak
magnetic field locally supercritical clumps are assembled due to
sufficient flux loss.

An interesting point is also the field morphology. In the weak
magnetic field case the field structure in the dense regions is
clearly separated from the large scale magnetic field, whereas in the
strong field case the field morphology is almost unaffected compared
to the initial configuration (see the blue stream lines of
Fig.~\ref{fig:mc-form-tilted}). From an observational point of view,
the field structure and its dynamical importance within molecular
clouds is still debated. On the one hand, some multi-scale
polarisation data indicate that magnetic fields in GMCs are
essentially just dragged in from larger scales and are dynamically
important \cite{Li11, Li11Nature, LiHB15, PLANCK-2015-XXXV}. On the
other hand, Zeeman measurements of individual cloud cores together
with analyses of numerical simulations indicate rather weak fields
that might not be dynamically important \cite{Crutcher09,
  Padoan10}. With our subsequent studies on cloud formation on cloud
scales including a more detailed modelling of ambipolar diffusion we
hope to clarify this issue.

\bef
\centering
  \includegraphics[width=\figthree@scaling\linewidth,angle=-90]{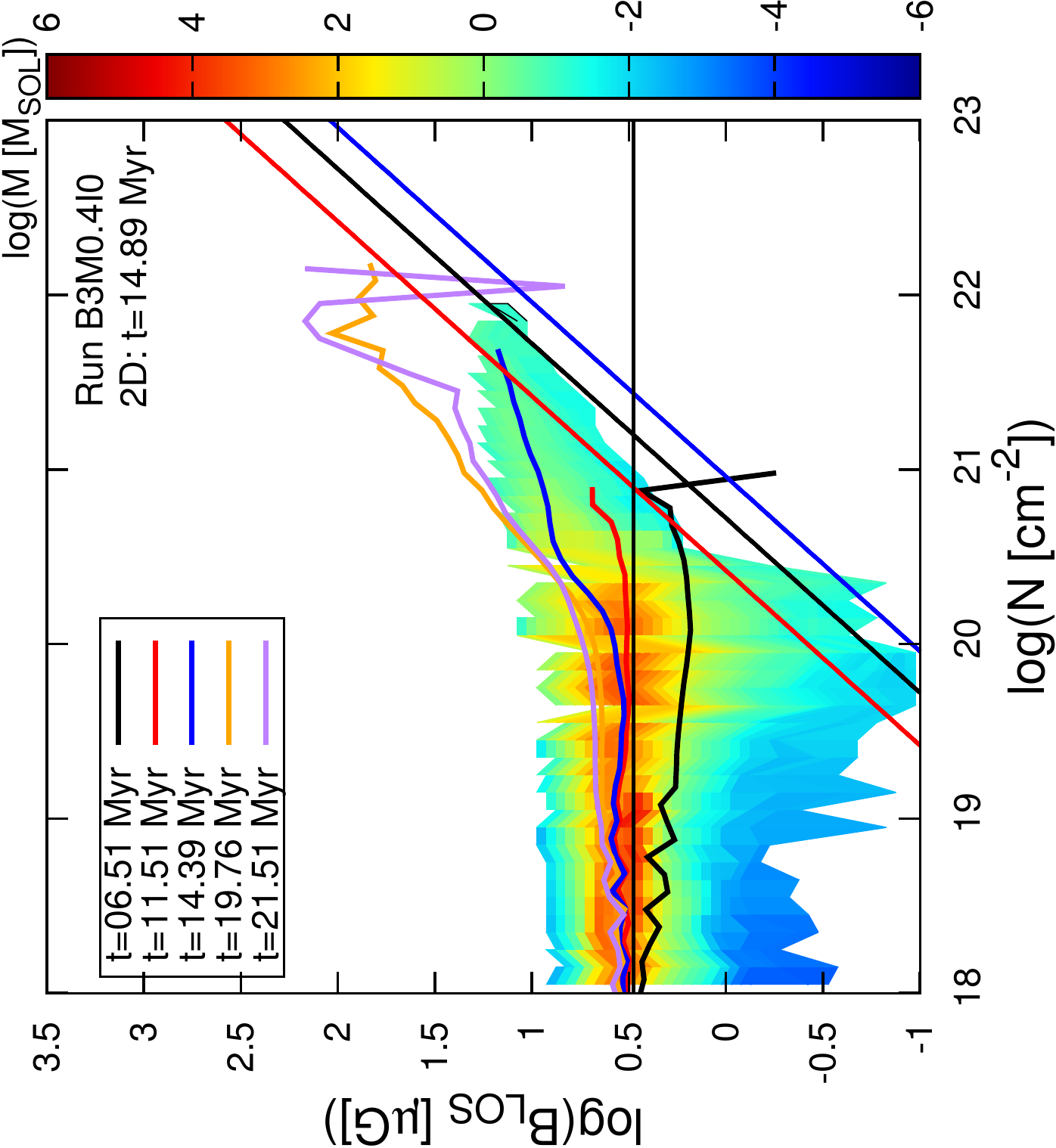}\,\includegraphics[width=\figthree@scaling\linewidth,angle=-90]{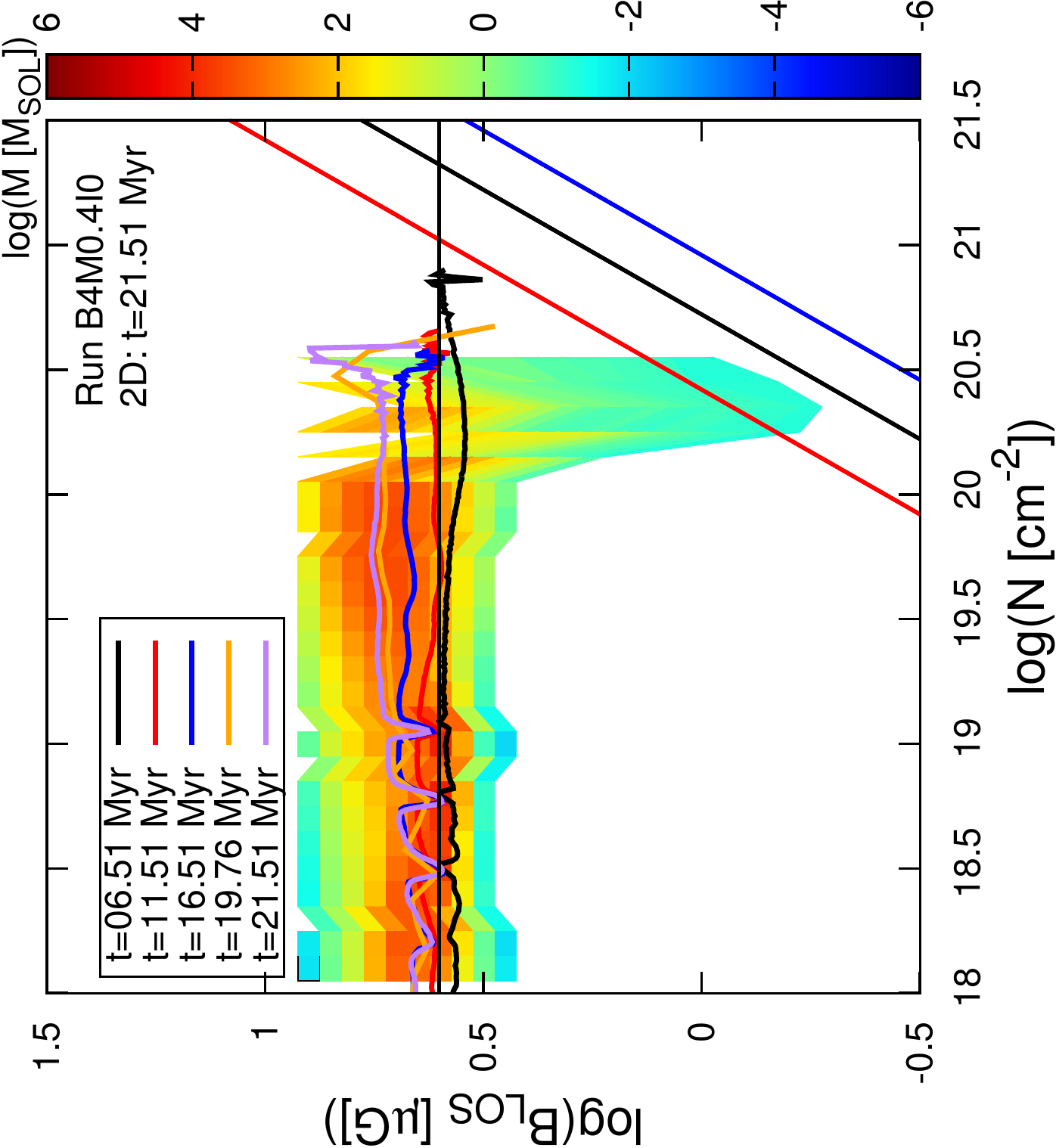}\,\includegraphics[width=\figthree@scaling\linewidth,angle=-90]{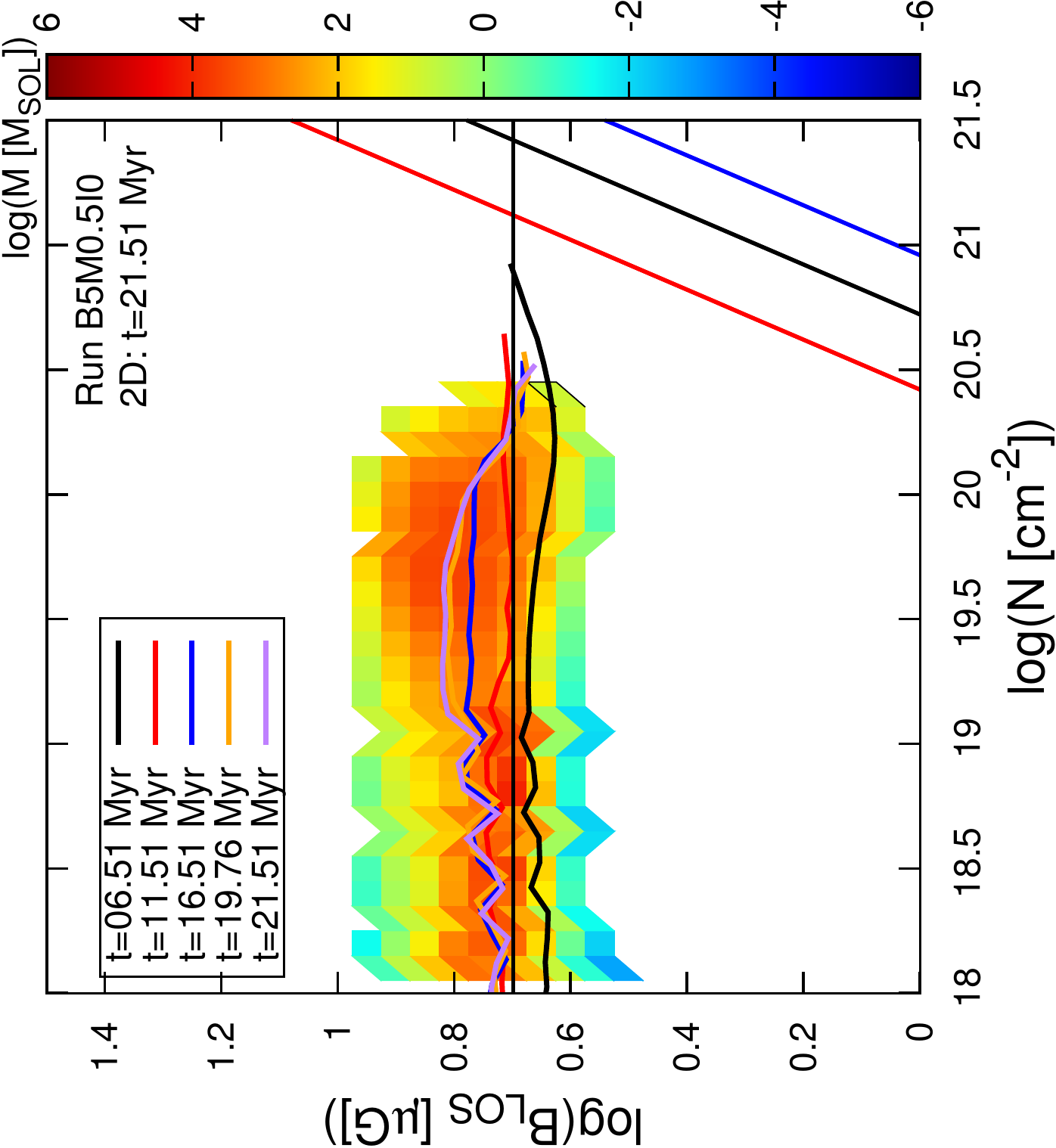}\\
  \includegraphics[width=\figthree@scaling\linewidth,angle=-90]{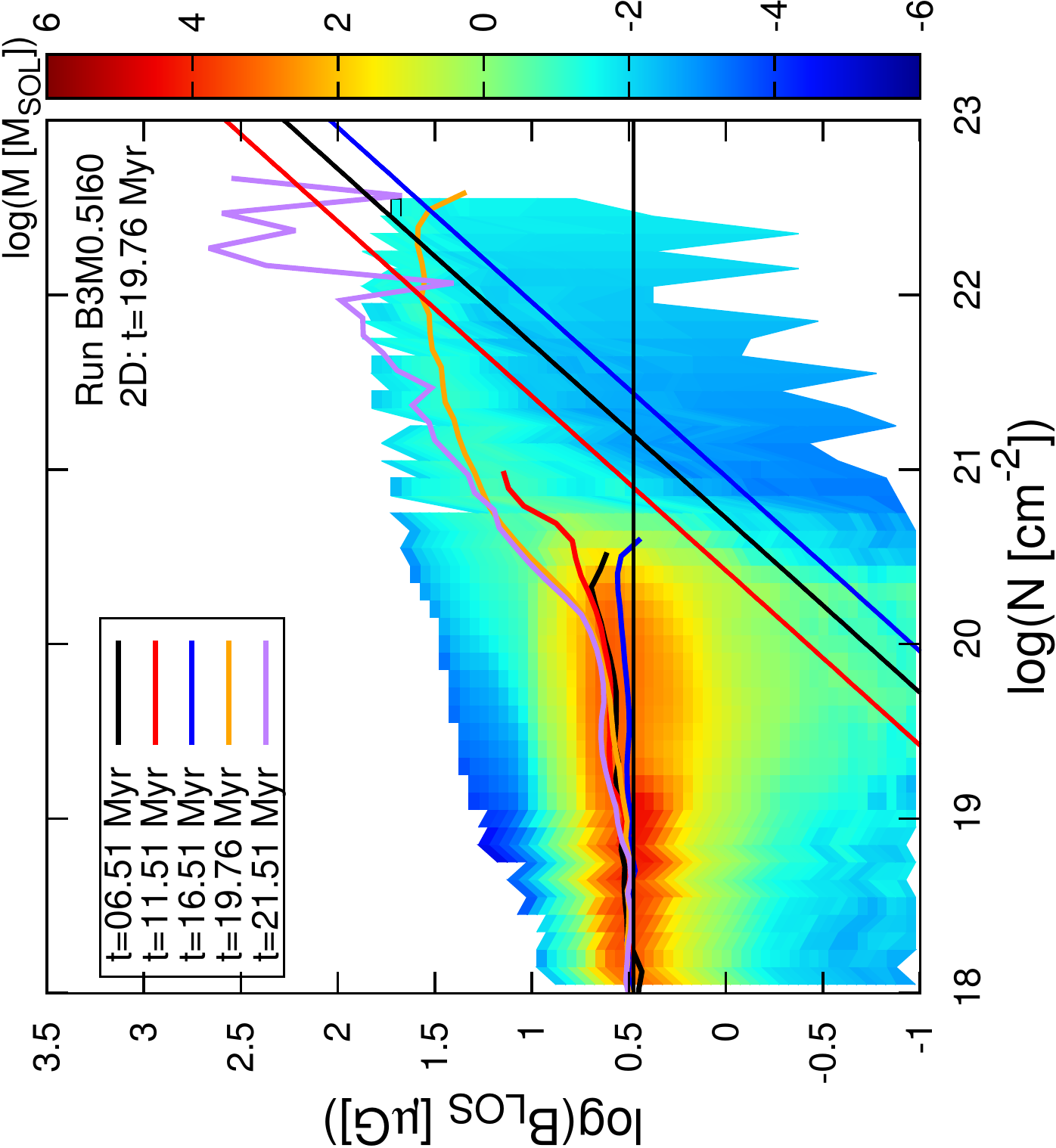}\,\includegraphics[width=\figthree@scaling\linewidth,angle=-90]{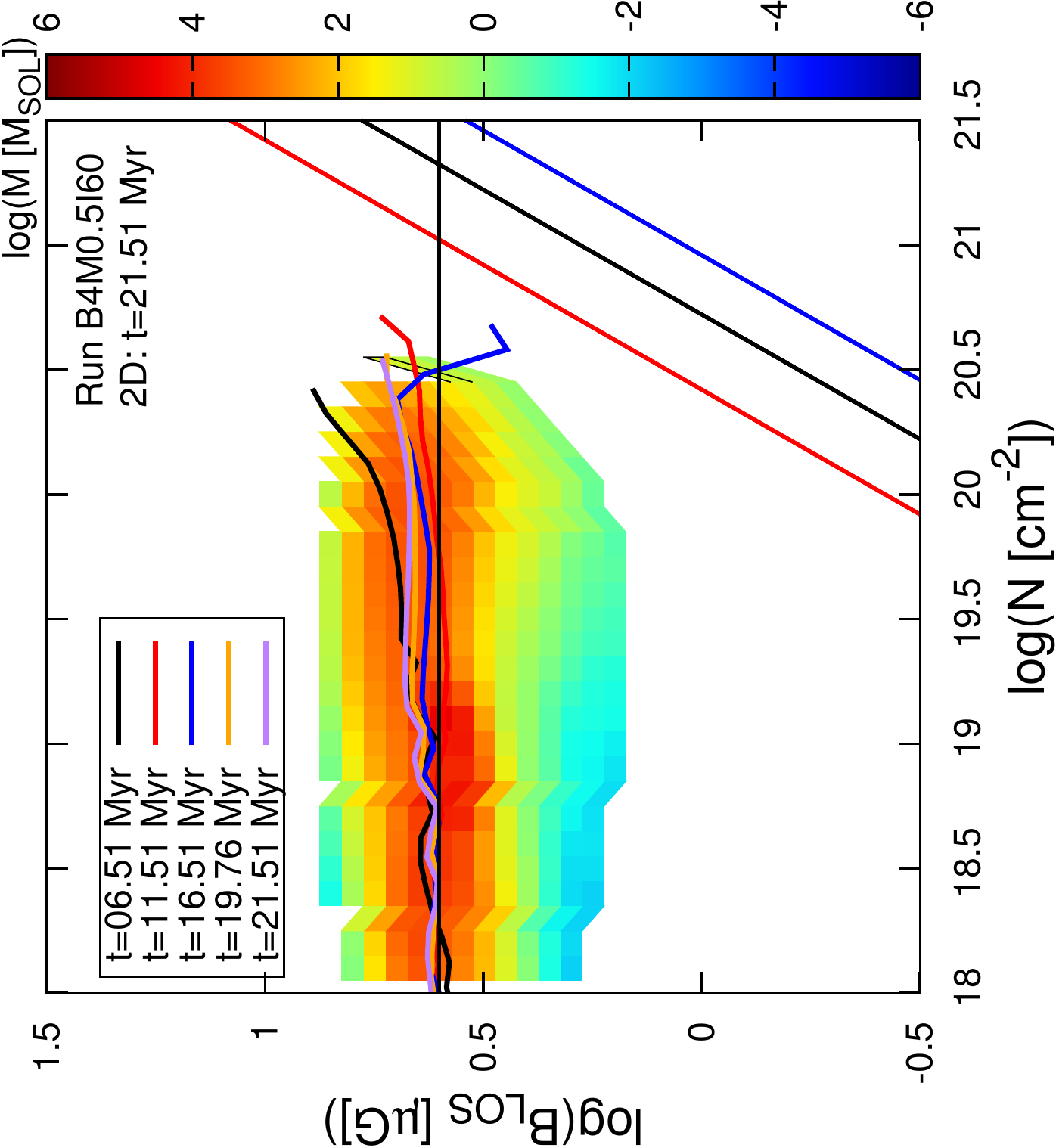}\,\includegraphics[width=\figthree@scaling\linewidth,angle=-90]{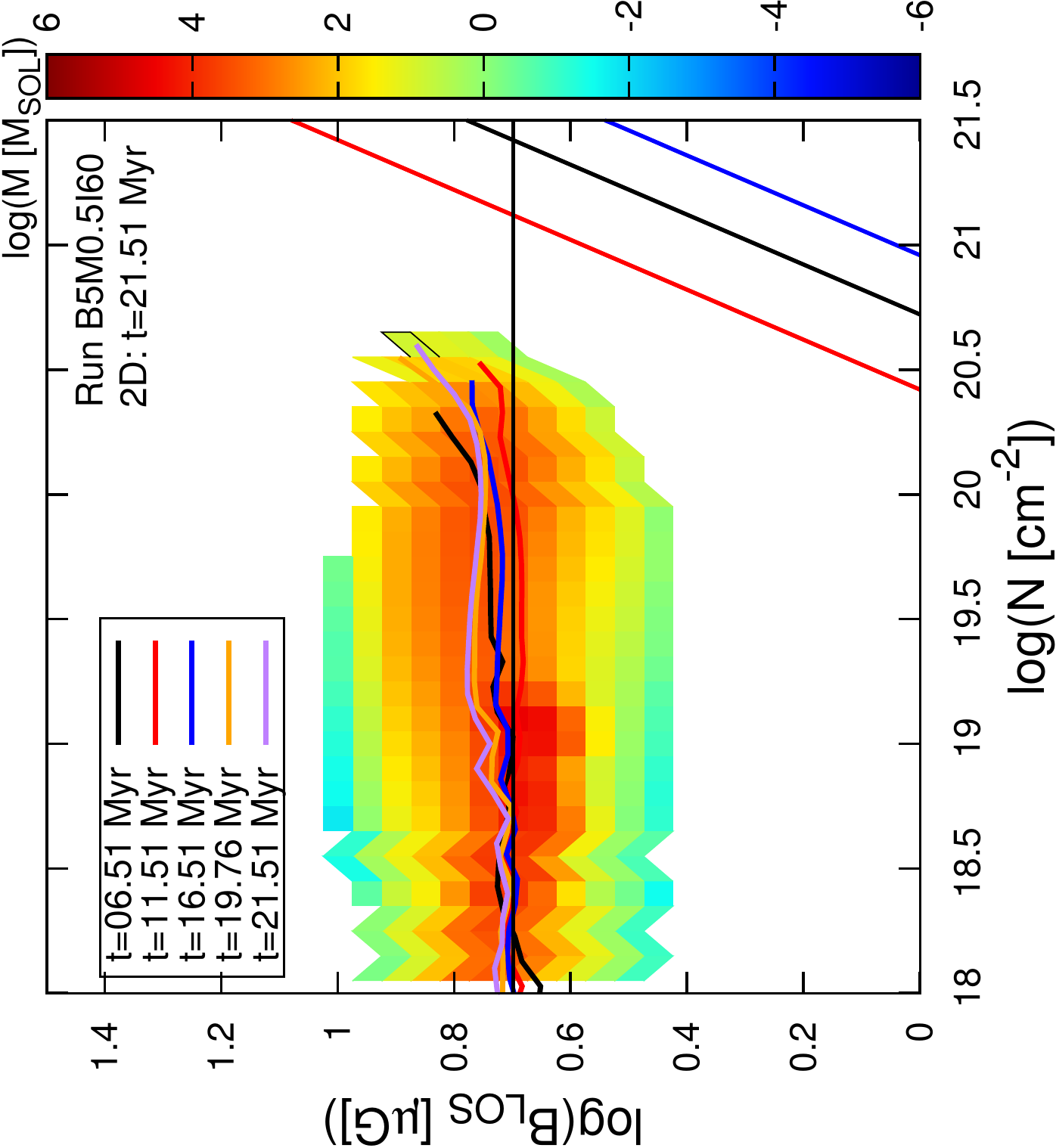}
  \caption{Results from colliding flow simulations with various
    different initial conditions. Shown are histograms of the
    line-of-sight field strength $B_{\sm{LOS}}$ as function of the
      column density $N$. From left to right:
      $B = 3\,\muG$,$B = 4\,\muG$ and $B = 5\,\muG$,
      respectively. Top:
      $\Phi=0^{\circ}$,
%$\Phi=30^{\circ}$,
      Bottom: $\Phi=60^{\circ}$. Different line colours denote
      different times. Also shown are the criticality condition
      (red line\cite{Crutcher10,Crutcher12}), corrected for
      projection effects (black line\cite{Shu99}), and assuming
      equipartition of turbulent and magnetic fields (blue
      line\cite{McKee93}). Colour coded is the mass distribution within
      this two parameter space. [from \KB\cite{Koertgen15}]}
\label{fig:BN}
\eef

In Fig.~\ref{fig:BN} we quantify the main results by means of
histograms in the $N$-$B$-plane from our colliding flow studies for
various initial conditions. Only in the {\it initially} marginally
subcritical case ($B = 3\,\muG$) we observe signs of star formation
within supercritical cores\footnote{If we assume
  $\mu_{\sm{crit}} \approx 0.13/\sqrt{G}$ for spherical cores
  \cite{Mouschovias76} we get
  $\mu/\mu_{\sm{crit}} = 0.97\,(3\,\muG/B)$ for our head-on colliding
  flow configurations.}. For slightly stronger initial magnetic fields
($B \simge 4\,\muG$) no supercritical cloud cores are generated, hence
there is no star formation activity, regardless whether ambipolar diffusion is
active or the flows collide with an oblique angle. Nevertheless the
results of those simulation show the observed behaviour in the low
column regime ($N \simle 10^{21}\,\cm^{-2}$), where gas assembles
along field lines without changing the field strength by much (see
also the latest analysis from PLANCK observations of individual
molecular clouds)\cite{PLANCK-2015-XXXV}. Only within supercritical,
self-gravitating cores the magnetic field gets enhanced by compression
due to flux freezing. Furthermore, we observe that star formation is
immediately initiated, when the gas becomes supercritical promoting a
picture of ``rapid'' star formation\cite{Hartmann01}.

\section{Conclusions}

Here, we summarised our recent results from MHD simulations of
colliding flows with varying initial conditions on the possible
formation of supercritical cloud cores from subcritical initial
conditions.  Although, dense clouds are easily formed within colliding
flow scenarios due to thermal instability, the generation of
supercritical clumps are largely determined by the initial conditions.
Furthermore, increasing initial turbulence lead to lower masses of the
cores and clumps because the HI streams become less
coherent. Otherwise, increasing magnetic field strength lead to more
massive molecular clouds, which nevertheless do not become
supercritical. Oblique flows still lead to cloud cores with masses
comparable to what has been observed recently. But starting with
subcritical HI flows, in no case the magnetic flux loss is sufficient
to allow the build--up of supercritical cloud cores. Generally,
increasing inclination of the flows lead to increasing diffusivity of
the magnetic field. Again, regardless of the variation of the
inclination, no tendency for faster accumulation of gas or faster
transition to thermally dominated regions was seen in our simulations.

We therefore stress the role of magnetic fields in the context of
molecular cloud and star formation. We point out the complete lack of
supercritical regions for realistic initial field strengths. From the
observational side, HI clouds may be supercritical as a whole, but
their observed, dense subregions be subcritical.

Hence, the question remains, how magnetically {\em supercritcal} cloud
cores are formed?

\section*{Acknowledgments}

{\small RB acknowledges funding by the DFG via the Emmy-Noether grant
  BA 3706/1-1, the ISM-SPP 1573 grants BA 3706/3-1 and BA 3706/3-2, as
  well as for the grant BA 3706/4-1.  The software used in this work
  was in part developed by the DOE--supported ASC/Alliance Center for
  Astrophysical Thermonuclear Flashes at the University of Chicago.
  The authors gratefully acknowledge the Gauss Centre for
  Supercomputing (GCS) for providing computing time through the John
  von Neumann Institute for Computing (NIC) on the GCS share of the
  supercomputer JUQUEEN\cite{JUQUEEN} at J\"ulich Supercomputing Centre
  (JSC). GCS is the alliance of the three national supercomputing
  centres HLRS (Universität Stuttgart), JSC (Forschungszentrum
  J\"ulich), and LRZ (Bayerische Akademie der Wissenschaften), funded by
  the German Federal Ministry of Education and Research (BMBF) and the
  German State Ministries for Research of Baden-W\"urttemberg (MWK),
  Bayern (StMWFK) and Nordrhein-Westfalen (MIWF).}

%The presented simulations were performed at the HLRN--III and the . }
%BK and RB thank Enrique V\'{a}zquez--Semadeni for useful
%discussions. BK acknowledges hospitality at Centro de
%Radioastronom\'{i}a y Astrof\'{i}sica, Universidad Nacional
%Aut\'{o}noma M\'{e}xico,  during the initial stages of this study. 

%\bibliographystyle{naturemag}
%\bibliography{astro.bib}

\begin{thebibliography}{10}
\expandafter\ifx\csname url\endcsname\relax
  \def\url#1{\texttt{#1}}\fi
\expandafter\ifx\csname urlprefix\endcsname\relax\def\urlprefix{URL }\fi
\providecommand{\bibinfo}[2]{#2}
\providecommand{\eprint}[2][]{\url{#2}}

\bibitem{WilliamsPPIV00}
\bibinfo{author}{{Williams}, J.~P.}, \bibinfo{author}{{Blitz}, L.} \&
  \bibinfo{author}{{McKee}, C.~F.}
\newblock \bibinfo{title}{{The Structure and Evolution of Molecular Clouds:
  from Clumps to Cores to the IMF}}.
\newblock \emph{\bibinfo{journal}{Protostars and Planets IV}}
  \bibinfo{pages}{97} (\bibinfo{year}{2000}).
\newblock \eprint{astro-ph/9902246}.

\bibitem{Blitz07}
\bibinfo{author}{{Blitz}, L.} \emph{et~al.}
\newblock \bibinfo{title}{{Giant Molecular Clouds in Local Group Galaxies}}.
\newblock \emph{\bibinfo{journal}{Protostars and Planets V}}
  \bibinfo{pages}{81--96} (\bibinfo{year}{2007}).
\newblock \eprint{arXiv:astro-ph/0602600}.

\bibitem{Shu87}
\bibinfo{author}{{Shu}, F.~H.}, \bibinfo{author}{{Adams}, F.~C.} \&
  \bibinfo{author}{{Lizano}, S.}
\newblock \bibinfo{title}{{Star formation in molecular clouds - Observation and
  theory}}.
\newblock \emph{\bibinfo{journal}{\araa}} \textbf{\bibinfo{volume}{25}},
  \bibinfo{pages}{23--81} (\bibinfo{year}{1987}).

\bibitem{MacLow04}
\bibinfo{author}{{Mac Low}, M.-M.} \& \bibinfo{author}{{Klessen}, R.~S.}
\newblock \bibinfo{title}{{Control of star formation by supersonic
  turbulence}}.
\newblock \emph{\bibinfo{journal}{Reviews of Modern Physics}}
  \textbf{\bibinfo{volume}{76}}, \bibinfo{pages}{125--194}
  (\bibinfo{year}{2004}).

\bibitem{AndrePPVI14}
\bibinfo{author}{{Andr{\'e}}, P.} \emph{et~al.}
\newblock \bibinfo{title}{{From Filamentary Networks to Dense Cores in
  Molecular Clouds: Toward a New Paradigm for Star Formation}}.
\newblock \emph{\bibinfo{journal}{Protostars and Planets VI}}
  \bibinfo{pages}{27--51} (\bibinfo{year}{2014}).
\newblock \eprint{1312.6232}.

\bibitem{Beck12}
\bibinfo{author}{{Beck}, R.}
\newblock \bibinfo{title}{{Magnetic Fields in Galaxies}}.
\newblock \emph{\bibinfo{journal}{\ssr}} \textbf{\bibinfo{volume}{166}},
  \bibinfo{pages}{215--230} (\bibinfo{year}{2012}).

\bibitem{Crutcher12}
\bibinfo{author}{{Crutcher}, R.~M.}
\newblock \bibinfo{title}{{Magnetic Fields in Molecular Clouds}}.
\newblock \emph{\bibinfo{journal}{\araa}} \textbf{\bibinfo{volume}{50}},
  \bibinfo{pages}{29--63} (\bibinfo{year}{2012}).

\bibitem{Heiles12}
\bibinfo{author}{{Heiles}, C.} \& \bibinfo{author}{{Haverkorn}, M.}
\newblock \bibinfo{title}{{Magnetic Fields in the Multiphase Interstellar
  Medium}}.
\newblock \emph{\bibinfo{journal}{\ssr}} \textbf{\bibinfo{volume}{166}},
  \bibinfo{pages}{293--305} (\bibinfo{year}{2012}).

\bibitem{Beck13}
\bibinfo{author}{{Beck}, R.} \& \bibinfo{author}{{Wielebinski}, R.}
\newblock \emph{\bibinfo{title}{{Magnetic Fields in Galaxies}}},
  \bibinfo{pages}{641} (\bibinfo{publisher}{Springer Science+Business Media
  Dordrecht}, \bibinfo{year}{2013}).

\bibitem{LiHBPPVI14}
\bibinfo{author}{{Li}, H.-B.} \emph{et~al.}
\newblock \bibinfo{title}{{The Link Between Magnetic Fields and Cloud/Star
  Formation}}.
\newblock \emph{\bibinfo{journal}{Protostars and Planets VI}}
  \bibinfo{pages}{101--123} (\bibinfo{year}{2014}).
\newblock \eprint{1404.2024}.

\bibitem{LiPPVI14}
\bibinfo{author}{{Li}, Z.-Y.} \emph{et~al.}
\newblock \bibinfo{title}{{The Earliest Stages of Star and Planet Formation:
  Core Collapse, and the Formation of Disks and Outflows}}.
\newblock \emph{\bibinfo{journal}{Protostars and Planets VI}}
  \bibinfo{pages}{173--194} (\bibinfo{year}{2014}).
\newblock \eprint{1401.2219}.

\bibitem{Li11Nature}
\bibinfo{author}{{Li}, H.-B.} \& \bibinfo{author}{{Henning}, T.}
\newblock \bibinfo{title}{{The alignment of molecular cloud magnetic fields
  with the spiral arms in M33}}.
\newblock \emph{\bibinfo{journal}{\nat}} \textbf{\bibinfo{volume}{479}},
  \bibinfo{pages}{499--501} (\bibinfo{year}{2011}).
\newblock \eprint{1111.2745}.

\bibitem{LiHB15}
\bibinfo{author}{{Li}, H.-B.} \emph{et~al.}
\newblock \bibinfo{title}{{Self-similar fragmentation regulated by magnetic
  fields in a region forming massive stars}}.
\newblock \emph{\bibinfo{journal}{\nat}} \textbf{\bibinfo{volume}{520}},
  \bibinfo{pages}{518--521} (\bibinfo{year}{2015}).

\bibitem{Pillai15}
\bibinfo{author}{{Pillai}, T.} \emph{et~al.}
\newblock \bibinfo{title}{{Magnetic Fields in High-mass Infrared Dark Clouds}}.
\newblock \emph{\bibinfo{journal}{\apj}} \textbf{\bibinfo{volume}{799}},
  \bibinfo{pages}{74} (\bibinfo{year}{2015}).
\newblock \eprint{1410.7390}.

\bibitem{Franco15}
\bibinfo{author}{{Franco}, G.~A.~P.} \& \bibinfo{author}{{Alves}, F.~O.}
\newblock \bibinfo{title}{{Tracing the magnetic field morphology of the Lupus I
  molecular cloud}}.
\newblock \emph{\bibinfo{journal}{ArXiv 1504.08222, accepted for publication in
  \apj}}  (\bibinfo{year}{2015}).
\newblock \eprint{1504.08222}.

\bibitem{Heiles05}
\bibinfo{author}{{Heiles}, C.} \& \bibinfo{author}{{Troland}, T.~H.}
\newblock \bibinfo{title}{{The Millennium Arecibo 21 Centimeter Absorption-Line
  Survey. IV. Statistics of Magnetic Field, Column Density, and Turbulence}}.
\newblock \emph{\bibinfo{journal}{\apj}} \textbf{\bibinfo{volume}{624}},
  \bibinfo{pages}{773--793} (\bibinfo{year}{2005}).
\newblock \eprint{arXiv:astro-ph/0501482}.

\bibitem{Mestel56}
\bibinfo{author}{{Mestel}, L.} \& \bibinfo{author}{{Spitzer}, L., Jr.}
\newblock \bibinfo{title}{{Star formation in magnetic dust clouds}}.
\newblock \emph{\bibinfo{journal}{\mnras}} \textbf{\bibinfo{volume}{116}},
  \bibinfo{pages}{503} (\bibinfo{year}{1956}).

\bibitem{Mouschovias99}
\bibinfo{author}{{Mouschovias}, T.~C.} \& \bibinfo{author}{{Ciolek}, G.~E.}
\newblock \bibinfo{title}{{Magnetic Fields and Star Formation: A Theory
  Reaching Adulthood}}.
\newblock In \bibinfo{editor}{{C.~J.~Lada \& N.~D.~Kylafis}} (ed.)
  \emph{\bibinfo{booktitle}{NATO ASIC Proc. 540: The Origin of Stars and
  Planetary Systems}}, \bibinfo{pages}{305--+} (\bibinfo{year}{1999}).

\bibitem{Myers86}
\bibinfo{author}{{Myers}, P.~C.} \emph{et~al.}
\newblock \bibinfo{title}{{Molecular clouds and star formation in the inner
  galaxy - A comparison of CO, H II, and far-infrared surveys}}.
\newblock \emph{\bibinfo{journal}{\apj}} \textbf{\bibinfo{volume}{301}},
  \bibinfo{pages}{398--422} (\bibinfo{year}{1986}).

\bibitem{Evans09}
\bibinfo{author}{{Evans}, N.~J., II} \emph{et~al.}
\newblock \bibinfo{title}{{The Spitzer c2d Legacy Results: Star-Formation Rates
  and Efficiencies; Evolution and Lifetimes}}.
\newblock \emph{\bibinfo{journal}{\apjs}} \textbf{\bibinfo{volume}{181}},
  \bibinfo{pages}{321--350} (\bibinfo{year}{2009}).
\newblock \eprint{0811.1059}.

\bibitem{Crutcher09}
\bibinfo{author}{{Crutcher}, R.~M.}, \bibinfo{author}{{Hakobian}, N.} \&
  \bibinfo{author}{{Troland}, T.~H.}
\newblock \bibinfo{title}{{Testing Magnetic Star Formation Theory}}.
\newblock \emph{\bibinfo{journal}{\apj}} \textbf{\bibinfo{volume}{692}},
  \bibinfo{pages}{844--855} (\bibinfo{year}{2009}).
\newblock \eprint{0807.2862}.

\bibitem{Bertram12}
\bibinfo{author}{{Bertram}, E.}, \bibinfo{author}{{Federrath}, C.},
  \bibinfo{author}{{Banerjee}, R.} \& \bibinfo{author}{{Klessen}, R.~S.}
\newblock \bibinfo{title}{{Statistical analysis of the mass-to-flux ratio in
  turbulent cores: effects of magnetic field reversals and dynamo
  amplification}}.
\newblock \emph{\bibinfo{journal}{\mnras}} \textbf{\bibinfo{volume}{420}},
  \bibinfo{pages}{3163--3173} (\bibinfo{year}{2012}).
\newblock \eprint{1111.5539}.

\bibitem{Zuckerman74}
\bibinfo{author}{{Zuckerman}, B.} \& \bibinfo{author}{{Evans}, N.~J., II}.
\newblock \bibinfo{title}{{Models of massive molecular clouds}}.
\newblock \emph{\bibinfo{journal}{\apjl}} \textbf{\bibinfo{volume}{192}},
  \bibinfo{pages}{L149--L152} (\bibinfo{year}{1974}).

\bibitem{Larson81}
\bibinfo{author}{{Larson}, R.~B.}
\newblock \bibinfo{title}{{Turbulence and star formation in molecular clouds}}.
\newblock \emph{\bibinfo{journal}{\mnras}} \textbf{\bibinfo{volume}{194}},
  \bibinfo{pages}{809--826} (\bibinfo{year}{1981}).

\bibitem{Solomon87}
\bibinfo{author}{{Solomon}, P.~M.}, \bibinfo{author}{{Rivolo}, A.~R.},
  \bibinfo{author}{{Barrett}, J.} \& \bibinfo{author}{{Yahil}, A.}
\newblock \bibinfo{title}{{Mass, luminosity, and line width relations of
  Galactic molecular clouds}}.
\newblock \emph{\bibinfo{journal}{\apj}} \textbf{\bibinfo{volume}{319}},
  \bibinfo{pages}{730--741} (\bibinfo{year}{1987}).

\bibitem{Elmegreen04}
\bibinfo{author}{{Elmegreen}, B.~G.} \& \bibinfo{author}{{Scalo}, J.}
\newblock \bibinfo{title}{{Interstellar Turbulence I: Observations and
  Processes}}.
\newblock \emph{\bibinfo{journal}{\araa}} \textbf{\bibinfo{volume}{42}},
  \bibinfo{pages}{211--273} (\bibinfo{year}{2004}).

\bibitem{Ballesteros07}
\bibinfo{author}{{Ballesteros-Paredes}, J.}, \bibinfo{author}{{Klessen},
  R.~S.}, \bibinfo{author}{{Mac Low}, M.-M.} \&
  \bibinfo{author}{{Vazquez-Semadeni}, E.}
\newblock \bibinfo{title}{{Molecular Cloud Turbulence and Star Formation}}.
\newblock In \bibinfo{editor}{{Reipurth}, B.}, \bibinfo{editor}{{Jewitt}, D.}
  \& \bibinfo{editor}{{Keil}, K.} (eds.) \emph{\bibinfo{booktitle}{Protostars
  and Planets V}}, \bibinfo{pages}{63--80} (\bibinfo{year}{2007}).

\bibitem{DobbsPPVI14}
\bibinfo{author}{{Dobbs}, C.~L.} \emph{et~al.}
\newblock \bibinfo{title}{{Formation of Molecular Clouds and Global Conditions
  for Star Formation}}.
\newblock \emph{\bibinfo{journal}{Protostars and Planets VI}}
  \bibinfo{pages}{3--26} (\bibinfo{year}{2014}).
\newblock \eprint{1312.3223}.

\bibitem{PadoanPPVI14}
\bibinfo{author}{{Padoan}, P.} \emph{et~al.}
\newblock \bibinfo{title}{{The Star Formation Rate of Molecular Clouds}}.
\newblock \emph{\bibinfo{journal}{Protostars and Planets VI}}
  \bibinfo{pages}{77--100} (\bibinfo{year}{2014}).
\newblock \eprint{1312.5365}.

\bibitem{Crutcher10}
\bibinfo{author}{{Crutcher}, R.~M.}, \bibinfo{author}{{Hakobian}, N.} \&
  \bibinfo{author}{{Troland}, T.~H.}
\newblock \bibinfo{title}{{Self-consistent analysis of OH Zeeman
  observations}}.
\newblock \emph{\bibinfo{journal}{\mnras}} \textbf{\bibinfo{volume}{402}},
  \bibinfo{pages}{L64--L66} (\bibinfo{year}{2010}).
\newblock \eprint{0912.3024}.

\bibitem{Crutcher10b}
\bibinfo{author}{{Crutcher}, R.~M.}
\newblock \bibinfo{title}{{Role of Magnetic Fields in Star Formation}}.
\newblock \emph{\bibinfo{journal}{Highlights of Astronomy}}
  \textbf{\bibinfo{volume}{15}}, \bibinfo{pages}{438--439}
  (\bibinfo{year}{2010}).

\bibitem{Fatuzzo02}
\bibinfo{author}{{Fatuzzo}, M.} \& \bibinfo{author}{{Adams}, F.~C.}
\newblock \bibinfo{title}{{Enhancement of Ambipolar Diffusion Rates through
  Field Fluctuations}}.
\newblock In \emph{\bibinfo{booktitle}{American Astronomical Society Meeting
  Abstracts \#200}}, vol.~\bibinfo{volume}{34} of
  \emph{\bibinfo{series}{Bulletin of the American Astronomical Society}},
  \bibinfo{pages}{769} (\bibinfo{year}{2002}).

\bibitem{Heitsch04}
\bibinfo{author}{{Heitsch}, F.}, \bibinfo{author}{{Zweibel}, E.~G.},
  \bibinfo{author}{{Slyz}, A.~D.} \& \bibinfo{author}{{Devriendt}, J.~E.~G.}
\newblock \bibinfo{title}{{Magnetic Flux Transport in the ISM Through Turbulent
  Ambipolar Diffusion}}.
\newblock \emph{\bibinfo{journal}{\apss}} \textbf{\bibinfo{volume}{292}},
  \bibinfo{pages}{45--51} (\bibinfo{year}{2004}).

\bibitem{Kudoh07}
\bibinfo{author}{{Kudoh}, T.}, \bibinfo{author}{{Basu}, S.},
  \bibinfo{author}{{Ogata}, Y.} \& \bibinfo{author}{{Yabe}, T.}
\newblock \bibinfo{title}{{Three-dimensional simulations of molecular cloud
  fragmentation regulated by magnetic fields and ambipolar diffusion}}.
\newblock \emph{\bibinfo{journal}{\mnras}} \textbf{\bibinfo{volume}{380}},
  \bibinfo{pages}{499--505} (\bibinfo{year}{2007}).
\newblock \eprint{0706.2696}.

\bibitem{Kudoh11}
\bibinfo{author}{{Kudoh}, T.} \& \bibinfo{author}{{Basu}, S.}
\newblock \bibinfo{title}{{Formation of Collapsing Cores in Subcritical
  Magnetic Clouds: Three-dimensional Magnetohydrodynamic Simulations with
  Ambipolar Diffusion}}.
\newblock \emph{\bibinfo{journal}{\apj}} \textbf{\bibinfo{volume}{728}},
  \bibinfo{pages}{123} (\bibinfo{year}{2011}).
\newblock \eprint{1012.5707}.

\bibitem{Vishniac99}
\bibinfo{author}{{Vishniac}, E.~T.} \& \bibinfo{author}{{Lazarian}, A.}
\newblock \bibinfo{title}{{Reconnection in the Interstellar Medium}}.
\newblock \emph{\bibinfo{journal}{\apj}} \textbf{\bibinfo{volume}{511}},
  \bibinfo{pages}{193--203} (\bibinfo{year}{1999}).

\bibitem{Eyink13}
\bibinfo{author}{{Eyink}, G.} \emph{et~al.}
\newblock \bibinfo{title}{{Flux-freezing breakdown in high-conductivity
  magnetohydrodynamic turbulence}}.
\newblock \emph{\bibinfo{journal}{\nat}} \textbf{\bibinfo{volume}{497}},
  \bibinfo{pages}{466--469} (\bibinfo{year}{2013}).

\bibitem{McKee89}
\bibinfo{author}{{McKee}, C.~F.}
\newblock \bibinfo{title}{{Photoionization-regulated star formation and the
  structure of molecular clouds}}.
\newblock \emph{\bibinfo{journal}{\apj}} \textbf{\bibinfo{volume}{345}},
  \bibinfo{pages}{782--801} (\bibinfo{year}{1989}).

\bibitem{Myers88}
\bibinfo{author}{{Myers}, P.~C.} \& \bibinfo{author}{{Goodman}, A.~A.}
\newblock \bibinfo{title}{{Magnetic molecular clouds - Indirect evidence for
  magnetic support and ambipolar diffusion}}.
\newblock \emph{\bibinfo{journal}{\apj}} \textbf{\bibinfo{volume}{329}},
  \bibinfo{pages}{392--405} (\bibinfo{year}{1988}).

\bibitem{Bertoldi92}
\bibinfo{author}{{Bertoldi}, F.} \& \bibinfo{author}{{McKee}, C.~F.}
\newblock \bibinfo{title}{{Pressure-confined clumps in magnetized molecular
  clouds}}.
\newblock \emph{\bibinfo{journal}{\apj}} \textbf{\bibinfo{volume}{395}},
  \bibinfo{pages}{140--157} (\bibinfo{year}{1992}).

\bibitem{Crutcher99}
\bibinfo{author}{{Crutcher}, R.~M.}, \bibinfo{author}{{Troland}, T.~H.},
  \bibinfo{author}{{Lazareff}, B.}, \bibinfo{author}{{Paubert}, G.} \&
  \bibinfo{author}{{Kaz{\` e}s}, I.}
\newblock \bibinfo{title}{{Detection of the CN Zeeman Effect in Molecular
  Clouds}}.
\newblock \emph{\bibinfo{journal}{\apjl}} \textbf{\bibinfo{volume}{514}},
  \bibinfo{pages}{L121--L124} (\bibinfo{year}{1999}).

\bibitem{Bourke01}
\bibinfo{author}{{Bourke}, T.~L.}, \bibinfo{author}{{Myers}, P.~C.},
  \bibinfo{author}{{Robinson}, G.} \& \bibinfo{author}{{Hyland}, A.~R.}
\newblock \bibinfo{title}{{New OH Zeeman Measurements of Magnetic Field
  Strengths in Molecular Clouds}}.
\newblock \emph{\bibinfo{journal}{\apj}} \textbf{\bibinfo{volume}{554}},
  \bibinfo{pages}{916--932} (\bibinfo{year}{2001}).
\newblock \eprint{astro-ph/0102469}.

\bibitem{Crutcher03}
\bibinfo{author}{{Crutcher}, R.}, \bibinfo{author}{{Heiles}, C.} \&
  \bibinfo{author}{{Troland}, T.}
\newblock \bibinfo{title}{{Observations of Interstellar Magnetic Fields}}.
\newblock In \bibinfo{editor}{{Falgarone}, E.} \& \bibinfo{editor}{{Passot},
  T.} (eds.) \emph{\bibinfo{booktitle}{Turbulence and Magnetic Fields in
  Astrophysics}}, vol. \bibinfo{volume}{614} of \emph{\bibinfo{series}{Lecture
  Notes in Physics, Berlin Springer Verlag}}, \bibinfo{pages}{155--181}
  (\bibinfo{year}{2003}).

\bibitem{Troland08}
\bibinfo{author}{{Troland}, T.~H.} \& \bibinfo{author}{{Crutcher}, R.~M.}
\newblock \bibinfo{title}{{Magnetic Fields in Dark Cloud Cores: Arecibo OH
  Zeeman Observations}}.
\newblock \emph{\bibinfo{journal}{\apj}} \textbf{\bibinfo{volume}{680}},
  \bibinfo{pages}{457--465} (\bibinfo{year}{2008}).
\newblock \eprint{0802.2253}.

\bibitem{Vazquez11a}
\bibinfo{author}{{V{\'a}zquez-Semadeni}, E.} \emph{et~al.}
\newblock \bibinfo{title}{{Molecular cloud evolution - IV. Magnetic fields,
  ambipolar diffusion and the star formation efficiency}}.
\newblock \emph{\bibinfo{journal}{\mnras}} \textbf{\bibinfo{volume}{414}},
  \bibinfo{pages}{2511--2527} (\bibinfo{year}{2011}).
\newblock \eprint{1101.3384}.

\bibitem{Koertgen15}
\bibinfo{author}{{K{\"o}rtgen}, B.} \& \bibinfo{author}{{Banerjee}, R.}
\newblock \bibinfo{title}{{Impact of magnetic fields on molecular cloud
  formation and evolution}}.
\newblock \emph{\bibinfo{journal}{\mnras}} \textbf{\bibinfo{volume}{451}},
  \bibinfo{pages}{3340--3353} (\bibinfo{year}{2015}).
\newblock \eprint{1502.03306}.

\bibitem{Heitsch14}
\bibinfo{author}{{Heitsch}, F.} \& \bibinfo{author}{{Hartmann}, L.}
\newblock \bibinfo{title}{{Accretion and diffusion time-scales in sheets and
  filaments}}.
\newblock \emph{\bibinfo{journal}{\mnras}} \textbf{\bibinfo{volume}{443}},
  \bibinfo{pages}{230--240} (\bibinfo{year}{2014}).
\newblock \eprint{1406.2191}.

\bibitem{Passot95}
\bibinfo{author}{{Passot}, T.}, \bibinfo{author}{{Vazquez-Semadeni}, E.} \&
  \bibinfo{author}{{Pouquet}, A.}
\newblock \bibinfo{title}{{A Turbulent Model for the Interstellar Medium. II.
  Magnetic Fields and Rotation}}.
\newblock \emph{\bibinfo{journal}{\apj}} \textbf{\bibinfo{volume}{455}},
  \bibinfo{pages}{536--+} (\bibinfo{year}{1995}).
\newblock \eprint{arXiv:astro-ph/9601182}.

\bibitem{Hartmann01}
\bibinfo{author}{{Hartmann}, L.}, \bibinfo{author}{{Ballesteros-Paredes}, J.}
  \& \bibinfo{author}{{Bergin}, E.~A.}
\newblock \bibinfo{title}{{Rapid Formation of Molecular Clouds and Stars in the
  Solar Neighborhood}}.
\newblock \emph{\bibinfo{journal}{\apj}} \textbf{\bibinfo{volume}{562}},
  \bibinfo{pages}{852--868} (\bibinfo{year}{2001}).
\newblock \eprint{arXiv:astro-ph/0108023}.

\bibitem{Fletcher11}
\bibinfo{author}{{Fletcher}, A.}, \bibinfo{author}{{Beck}, R.},
  \bibinfo{author}{{Shukurov}, A.}, \bibinfo{author}{{Berkhuijsen}, E.~M.} \&
  \bibinfo{author}{{Horellou}, C.}
\newblock \bibinfo{title}{{Magnetic fields and spiral arms in the galaxy M51}}.
\newblock \emph{\bibinfo{journal}{\mnras}} \textbf{\bibinfo{volume}{412}},
  \bibinfo{pages}{2396--2416} (\bibinfo{year}{2011}).
\newblock \eprint{1001.5230}.

\bibitem{PLANCK-2015-XXXV}
\bibinfo{author}{{Planck Collaboration}} \emph{et~al.}
\newblock \bibinfo{title}{{Planck intermediate results. XXXV. Probing the role
  of the magnetic field in the formation of structure in molecular clouds}}.
\newblock \emph{\bibinfo{journal}{ArXiv e-prints: 1502.04123}}
  (\bibinfo{year}{2015}).
\newblock \eprint{1502.04123}.

\bibitem{FLASH00}
\bibinfo{author}{{Fryxell}, B.} \emph{et~al.}
\newblock \bibinfo{title}{{FLASH: An Adaptive Mesh Hydrodynamics Code for
  Modeling Astrophysical Thermonuclear Flashes}}.
\newblock \emph{\bibinfo{journal}{\apjs}} \textbf{\bibinfo{volume}{131}},
  \bibinfo{pages}{273--334} (\bibinfo{year}{2000}).

\bibitem{Dubey08}
\bibinfo{author}{{Dubey}, A.} \emph{et~al.}
\newblock \bibinfo{title}{{Challenges of Extreme Computing using the FLASH
  code}}.
\newblock In \bibinfo{editor}{{Pogorelov}, N.~V.}, \bibinfo{editor}{{Audit},
  E.} \& \bibinfo{editor}{{Zank}, G.~P.} (eds.)
  \emph{\bibinfo{booktitle}{Numerical Modeling of Space Plasma Flows}}, vol.
  \bibinfo{volume}{385} of \emph{\bibinfo{series}{Astronomical Society of the
  Pacific Conference Series}}, \bibinfo{pages}{145--+} (\bibinfo{year}{2008}).

\bibitem{Bouchut07}
\bibinfo{author}{{Bouchut}, F.}, \bibinfo{author}{{Klingenberg}, C.} \&
  \bibinfo{author}{{Waagan}, K.}
\newblock \bibinfo{title}{{A multiwave approximate Riemann solver for ideal MHD
  based on relaxation. I: theoretical framework}}.
\newblock \emph{\bibinfo{journal}{Numerische Mathematik}}
  \textbf{\bibinfo{volume}{108}}, \bibinfo{pages}{7--42}
  (\bibinfo{year}{2007}).

\bibitem{Bouchut10}
\bibinfo{author}{{Bouchut}, F.}, \bibinfo{author}{{Klingenberg}, C.} \&
  \bibinfo{author}{{Waagan}, K.}
\newblock \bibinfo{title}{{A multiwave approximate Riemann solver for ideal MHD
  based on relaxation II - Numerical implementation with 3 and 5 waves}}.
\newblock \emph{\bibinfo{journal}{Numerische Mathematik}}
  \textbf{\bibinfo{volume}{115}}, \bibinfo{pages}{647--679}
  (\bibinfo{year}{2010}).

\bibitem{Waagan10}
\bibinfo{author}{{Waagan}, K.}, \bibinfo{author}{{Federrath}, C.} \&
  \bibinfo{author}{{Klingenberg}, C.}
\newblock \bibinfo{title}{{A robust numerical code for compressible
  magnetohydrodynamics and its application to highly supersonic turbulence}}.
\newblock \emph{\bibinfo{journal}{\jcop, submitted}}  (\bibinfo{year}{2010}).

\bibitem{Duffin08}
\bibinfo{author}{{Duffin}, D.~F.} \& \bibinfo{author}{{Pudritz}, R.~E.}
\newblock \bibinfo{title}{{Simulating hydromagnetic processes in star
  formation: introducing ambipolar diffusion into an adaptive mesh refinement
  code}}.
\newblock \emph{\bibinfo{journal}{\mnras}} \textbf{\bibinfo{volume}{391}},
  \bibinfo{pages}{1659--1673} (\bibinfo{year}{2008}).
\newblock \eprint{0810.0299}.

\bibitem{Koyama02}
\bibinfo{author}{{Koyama}, H.} \& \bibinfo{author}{{Inutsuka}, S.-i.}
\newblock \bibinfo{title}{{An Origin of Supersonic Motions in Interstellar
  Clouds}}.
\newblock \emph{\bibinfo{journal}{\apjl}} \textbf{\bibinfo{volume}{564}},
  \bibinfo{pages}{L97--L100} (\bibinfo{year}{2002}).
\newblock \eprint{arXiv:astro-ph/0112420}.

\bibitem{Koyama00}
\bibinfo{author}{{Koyama}, H.} \& \bibinfo{author}{{Inutsuka}, S.-I.}
\newblock \bibinfo{title}{{Molecular Cloud Formation in Shock-compressed
  Layers}}.
\newblock \emph{\bibinfo{journal}{\apj}} \textbf{\bibinfo{volume}{532}},
  \bibinfo{pages}{980--993} (\bibinfo{year}{2000}).
\newblock \eprint{arXiv:astro-ph/9912509}.

\bibitem{Vazquez07}
\bibinfo{author}{{V{\'a}zquez-Semadeni}, E.} \emph{et~al.}
\newblock \bibinfo{title}{{Molecular Cloud Evolution. II. From Cloud Formation
  to the Early Stages of Star Formation in Decaying Conditions}}.
\newblock \emph{\bibinfo{journal}{\apj}} \textbf{\bibinfo{volume}{657}},
  \bibinfo{pages}{870--883} (\bibinfo{year}{2007}).
\newblock \eprint{astro-ph/0608375}.

\bibitem{Banerjee09a}
\bibinfo{author}{{Banerjee}, R.}, \bibinfo{author}{{V{\'a}zquez-Semadeni}, E.},
  \bibinfo{author}{{Hennebelle}, P.} \& \bibinfo{author}{{Klessen}, R.~S.}
\newblock \bibinfo{title}{{Clump morphology and evolution in MHD simulations of
  molecular cloud formation}}.
\newblock \emph{\bibinfo{journal}{\mnras}} \textbf{\bibinfo{volume}{398}},
  \bibinfo{pages}{1082--1092} (\bibinfo{year}{2009}).
\newblock \eprint{0808.0986}.

\bibitem{Federrath10}
\bibinfo{author}{{Federrath}, C.}, \bibinfo{author}{{Banerjee}, R.},
  \bibinfo{author}{{Clark}, P.~C.} \& \bibinfo{author}{{Klessen}, R.~S.}
\newblock \bibinfo{title}{{Modeling Collapse and Accretion in Turbulent Gas
  Clouds: Implementation and Comparison of Sink Particles in AMR and SPH}}.
\newblock \emph{\bibinfo{journal}{\apj}} \textbf{\bibinfo{volume}{713}},
  \bibinfo{pages}{269--290} (\bibinfo{year}{2010}).
\newblock \eprint{1001.4456}.

\bibitem{Truelove97}
\bibinfo{author}{{Truelove}, J.~K.} \emph{et~al.}
\newblock \bibinfo{title}{{The Jeans Condition: A New Constraint on Spatial
  Resolution in Simulations of Isothermal Self-gravitational Hydrodynamics}}.
\newblock \emph{\bibinfo{journal}{\apjl}} \textbf{\bibinfo{volume}{489}},
  \bibinfo{pages}{L179+} (\bibinfo{year}{1997}).

\bibitem{Vishniac94}
\bibinfo{author}{{Vishniac}, E.~T.}
\newblock \bibinfo{title}{{Nonlinear instabilities in shock-bounded slabs}}.
\newblock \emph{\bibinfo{journal}{\apj}} \textbf{\bibinfo{volume}{428}},
  \bibinfo{pages}{186--208} (\bibinfo{year}{1994}).
\newblock \eprint{astro-ph/9306025}.

\bibitem{Mouschovias76}
\bibinfo{author}{{Mouschovias}, T.~C.} \& \bibinfo{author}{{Spitzer}, L., Jr.}
\newblock \bibinfo{title}{{Note on the collapse of magnetic interstellar
  clouds}}.
\newblock \emph{\bibinfo{journal}{\apj}} \textbf{\bibinfo{volume}{210}},
  \bibinfo{pages}{326--+} (\bibinfo{year}{1976}).

\bibitem{Heitsch05}
\bibinfo{author}{{Heitsch}, F.}, \bibinfo{author}{{Burkert}, A.},
  \bibinfo{author}{{Hartmann}, L.~W.}, \bibinfo{author}{{Slyz}, A.~D.} \&
  \bibinfo{author}{{Devriendt}, J.~E.~G.}
\newblock \bibinfo{title}{{Formation of Structure in Molecular Clouds: A Case
  Study}}.
\newblock \emph{\bibinfo{journal}{\apjl}} \textbf{\bibinfo{volume}{633}},
  \bibinfo{pages}{L113--L116} (\bibinfo{year}{2005}).
\newblock \eprint{arXiv:astro-ph/0507567}.

\bibitem{Heitsch08b}
\bibinfo{author}{{Heitsch}, F.}, \bibinfo{author}{{Hartmann}, L.} \&
  \bibinfo{author}{{Burkert}, A.}
\newblock \bibinfo{title}{{Fragmentation of Shocked Flows: Gravity, Turbulence
  and Cooling}}.
\newblock \emph{\bibinfo{journal}{ArXiv e-prints}}
  \textbf{\bibinfo{volume}{805}} (\bibinfo{year}{2008}).
\newblock \eprint{0805.0801}.

\bibitem{Inoue08}
\bibinfo{author}{{Inoue}, T.} \& \bibinfo{author}{{Inutsuka}, S.-i.}
\newblock \bibinfo{title}{{Two-Fluid Magnetohydrodynamic Simulations of
  Converging H I Flows in the Interstellar Medium. I. Methodology and Basic
  Results}}.
\newblock \emph{\bibinfo{journal}{\apj}} \textbf{\bibinfo{volume}{687}},
  \bibinfo{pages}{303--310} (\bibinfo{year}{2008}).
\newblock \eprint{0801.0486}.

\bibitem{Field65}
\bibinfo{author}{{Field}, G.~B.}
\newblock \bibinfo{title}{{Thermal Instability.}}
\newblock \emph{\bibinfo{journal}{\apj}} \textbf{\bibinfo{volume}{142}},
  \bibinfo{pages}{531--+} (\bibinfo{year}{1965}).

\bibitem{Li11}
\bibinfo{author}{{Li}, H.-B.} \emph{et~al.}
\newblock \bibinfo{title}{{Evidence for dynamically important magnetic fields
  in molecular clouds}}.
\newblock \emph{\bibinfo{journal}{\mnras}} \textbf{\bibinfo{volume}{411}},
  \bibinfo{pages}{2067--2075} (\bibinfo{year}{2011}).
\newblock \eprint{1007.3312}.

\bibitem{Padoan10}
\bibinfo{author}{{Padoan}, P.} \emph{et~al.}
\newblock \bibinfo{title}{{MHD Turbulence In Star-Forming Clouds}}.
\newblock In \bibinfo{editor}{{Bertin}, G.}, \bibinfo{editor}{{de Luca}, F.},
  \bibinfo{editor}{{Lodato}, G.}, \bibinfo{editor}{{Pozzoli}, R.} \&
  \bibinfo{editor}{{Rom{\'e}}, M.} (eds.) \emph{\bibinfo{booktitle}{American
  Institute of Physics Conference Series}}, vol. \bibinfo{volume}{1242} of
  \emph{\bibinfo{series}{American Institute of Physics Conference Series}},
  \bibinfo{pages}{219--230} (\bibinfo{year}{2010}).

\bibitem{Shu99}
\bibinfo{author}{{Shu}, F.~H.}, \bibinfo{author}{{Allen}, A.},
  \bibinfo{author}{{Shang}, H.}, \bibinfo{author}{{Ostriker}, E.~C.} \&
  \bibinfo{author}{{Li}, Z.-Y.}
\newblock \bibinfo{title}{{Low-Mass Star Formation: Theory}}.
\newblock In \bibinfo{editor}{{Lada}, C.~J.} \& \bibinfo{editor}{{Kylafis},
  N.~D.} (eds.) \emph{\bibinfo{booktitle}{NATO Advanced Science Institutes
  (ASI) Series C}}, vol. \bibinfo{volume}{540} of \emph{\bibinfo{series}{NATO
  Advanced Science Institutes (ASI) Series C}}, \bibinfo{pages}{193}
  (\bibinfo{year}{1999}).

\bibitem{McKee93}
\bibinfo{author}{{McKee}, C.~F.}, \bibinfo{author}{{Zweibel}, E.~G.},
  \bibinfo{author}{{Goodman}, A.~A.} \& \bibinfo{author}{{Heiles}, C.}
\newblock \bibinfo{title}{{Magnetic Fields in Star-Forming Regions - Theory}}.
\newblock In \bibinfo{editor}{{Levy}, E.~H.} \& \bibinfo{editor}{{Lunine},
  J.~I.} (eds.) \emph{\bibinfo{booktitle}{Protostars and Planets III}},
  \bibinfo{pages}{327--+} (\bibinfo{year}{1993}).

\bibitem{JUQUEEN}
\bibinfo{author}{{J\"ulich Supercomputing Centre}}.
\newblock \bibinfo{title}{{JUQUEEN: IBM Blue Gene/Q Supercomputer System at the
  J\"ulich Supercomputing Centre}}.
\newblock \emph{\bibinfo{journal}{Journal of large-scale research facilities}}
  \textbf{\bibinfo{volume}{1}}, \bibinfo{pages}{A1} (\bibinfo{year}{2015}).

\end{thebibliography}

\end{document}